\documentclass[aps,prb,superscriptaddress,twocolumn,floatfix]{revtex4}
\usepackage{psfig}

\begin{document}

\title{
\parbox{30mm}{\fbox{\rule[1mm]{2mm}{-2mm}\Large\bf\sf PREPRINT}}
\hspace*{4mm}
\parbox{100mm}{\footnotesize\sf
Submitted for publication in {\it Optics Letters} (2004) }\hfill
%%%%%%%%%%%%%%%%
\\[5mm]
Observation of surface states in a truncated two-dimensional
photonic crystal slab}
\date{\today}

\author{Yurii~A.\ Vlasov}
\email[Corresponding author; e-mail: ]{yvlasov@us.ibm.com}
\affiliation{IBM T.~J.\ Watson Research Center, Yorktown Heights, NY
  10598, USA}

\author{Nikolaj\ Moll}
\affiliation{IBM Research, Zurich Research Laboratory,
R\"uschlikon, Switzerland}

\author{Sharee~J.\ McNab}
\affiliation{IBM T.~J.\ Watson Research Center, Yorktown Heights, NY
  10598, USA}

\begin{abstract}
The effect of lattice termination on the surface states in a
two-dimensional truncated photonic crystal slab is experimentally
studied in a high index contrast silicon-on-insulator system. A
single-mode silicon strip waveguide that is separated from the
photonic crystal by a trench of variable width is used to
evanescently couple to surface states in the surrounding lattice.
It is demonstrated that the dispersion of the surface states
depends strongly on the specific termination of the lattice.
\end{abstract}

\maketitle

Two-dimensional (2D) slab-type photonic crystals (PhC) are viewed
as a promising platform for dense integration of discrete optical
components into a photonic integrated circuit (PIC).  In almost
all of the optical devices based on 2D PhCs the periodicity of the
PhC lattice is locally disrupted to give rise to localized states
useful for the required functionality. However there is another
general class of localized states in PhCs - surface states
localized at the interface between the PhC and the surrounding
media \cite{ref1}, which can become increasingly important for
engineering dense PICs. As in any finite periodic system, the
surface states in the PhC originate from the abrupt termination of
the crystal's periodicity \cite{ref2}. Coupling of the waveguide
modes or cavity resonances to surface states in improperly
truncated PhC devices can disrupt light propagation and compromise
device performance. Surface states can also affect the impedance,
for example at the interface of a PhC waveguide and a strip
waveguide \cite{ref3} and therefore different truncations of the
lattice could be applied to improve impedance matching to reduce
coupling losses. Beaming and focusing has also been recently
demonstrated \cite{ref4,ref5} by changing the termination of the
PhC lattice and hence surface states.

\begin{figure}[tb]
\begin{center}
\leavevmode \psfig{figure=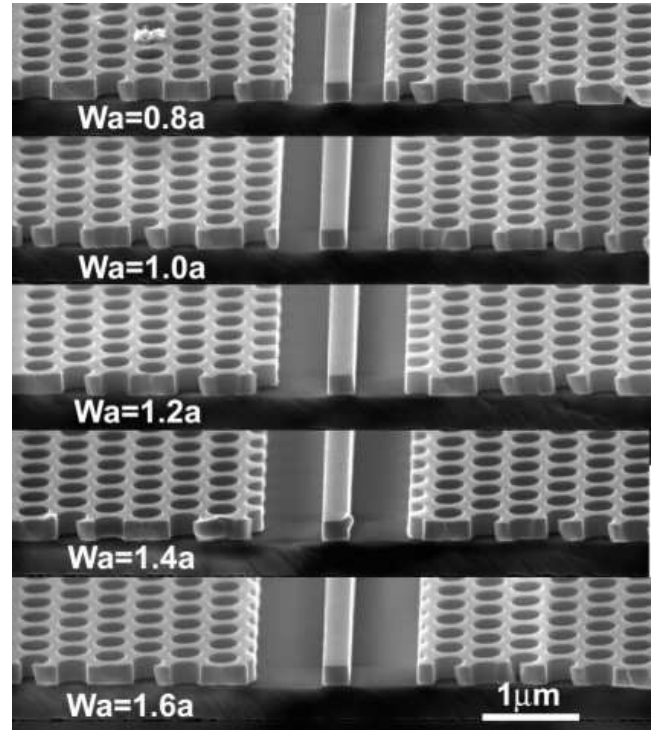,width=85mm}
\end{center}
\caption{Set of SEM images of the PhC DT waveguides with different
trench width ($W_a$) varying from 0.8$a$ to 1.6$a$ (top to
bottom).}
\label{fig1}
\end{figure}

In this paper we present experimental studies of surface states in
truncated 2D PhC slabs fabricated in silicon-on-insulator (SOI)
wafers. 200mm diameter SOI wafers with 220nm thick silicon device
layer on top of 2$\mu$m thick buried oxide (BOX) layer were
patterned with electron beam lithography and processed on a
standard CMOS fabrication line as reported elsewhere \cite{ref6}.
PhCs with a triangular lattice were defined by etching a periodic
array of circular holes down to the BOX layer. The hole diameter
$D$ was set to 306nm with a lattice constant $a$ of 437nm and a
silicon slab thickness of 0.5$a$ (220nm). Double-trench (DT) PhC
waveguides consisting of a conventional strip waveguide embedded
in a PhC lattice [7, 8], were explored as a model system for
probing surface states. The core of the waveguide is formed by
omitting one row of holes in the PhC lattice in the $\Gamma$-$M$
direction and etching two parallel trenches to define the strip
waveguide in the center. The width of the strip waveguide $W$
embedded in the PhC slab was chosen to be 0.6$a$ (263nm). The
double-trench design of the PhC waveguide allows the coupling
between the modes in the strip and in the PhC to be varied by
changing the width of the trench $W_a$ that separates the strip
from the surrounding PhC lattice. At the same time a different
$W_a$ implies a different truncation of the nearby holes of the
PhC lattice as shown in Fig.\ \ref{fig1}, thus providing a tool to
manipulate surface states.

Surface states are optically probed by inspecting the transmission
through the DT PhC waveguide with different truncations of the PhC
lattice adjacent to the waveguide channel. A pair of strip access
waveguides of the same width $W$ is used for coupling into and out
of the PhC waveguide. These narrow strip waveguides are
adiabatically tapered to a wider access waveguide with a final
width of 465nm.  For efficient coupling of light into and out of
the device-under-test (DUT) the access strip waveguides were
terminated by a pair of spot-size converters based on an inverted
taper design \cite{ref6}. The light from a broadband (1200-1700nm)
LED source is coupled to the DUT via a tapered and microlensed PM
fiber and, after transmission through the DUT, is collected by a
microlensed SM fiber and analyzed with an OSA. Further details of
the experimental set-up are reported elsewhere \cite{ref6}.

\begin{figure}[tb]
\begin{center}
\leavevmode \psfig{figure=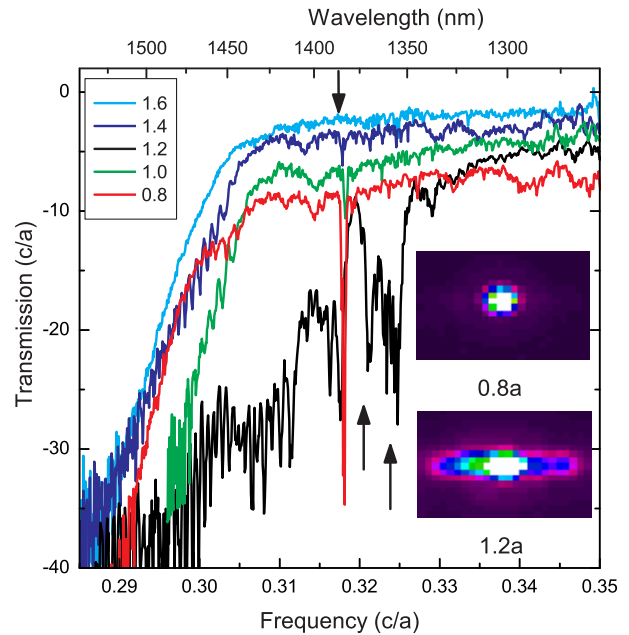,width=85mm}
\end{center}
\caption{Set of transmission spectra for TE-polarized light
through DT PhC waveguides with different trench widths. Spectra
were measured with a resolution of 0.5nm on waveguides 500$\mu$m
long. The cyan, blue, black, green and red spectra correspond to
nominal trench widths of 1.6$a$, 1.4$a$, 1.2$a$, 1.0$a$, and
0.8$a$ respectively. Spectra are shifted vertically with respect
to each other by 1dB for clarity. Inset: Field profiles of the
TE-mode in DT PhC waveguides with trench widths of 0.8$a$ (top)
and 1.2$a$ (bottom).The width of the field of view corresponds to
20$\mu$m.}
\label{fig2}
\end{figure}

Transmission spectra for the TE polarization measured on a series
of devices are shown in Fig.\ \ref{fig2} for different trench
widths $W_a$. The spectrum for the trench width of $W_a$=1.6$a$
can be used as a reference since the interaction with the PhC
modes is negligible for such a wide trench and the spectrum is
almost identical to that of the isolated strip waveguide. The
spectrum is characterized by a nearly flat band transmission over
a broad range of frequencies with a long-wavelength cut-off around
0.304$c/a$ (wavelength of 1440nm). With the decrease of the trench
width, and correspondingly the distance to the PhC, a narrow dip
appears at 0.317$c/a$ (wavelength of 1385nm). This dip was
previously identified as originating from a narrow stopband
appearing at the Brillouin-zone boundary \cite{ref8}. The
attenuation at the center of this dip gradually increases from
2.5dB to 26dB as the trench width is decreased from 1.4$a$ to
0.8$a$, and can be used as a measure of the magnitude of
interaction of the mode in the strip waveguide with the periodic
PhC lattice.

The spectrum for the trench width of 1.2$a$ is drastically
different from all other spectra in the set. Two new sharp bands
with strong attenuation (over 18dB) appear at frequencies
0.322$c/a$ and 0.325$c/a$. According to theoretical predictions in
Ref.7, the surface termination, which corresponds to $W_a$~
=1.2$a$, is favorable for moving the surface states to the center
of the photonic gap. Correspondingly these changes in the
transmission spectrum can be interpreted as coupling of the mode
in the strip waveguide to the surface states localized at the PhC
interface. This interpretation is also supported by the analysis
of the field pattern of the waveguiding mode shown in the inset of
Fig.\ \ref{fig2}. The field pattern of the propagating mode was
acquired with an IR camera through a 40X objective at the exit of
a cleaved 500$\mu$m long DT PhC waveguide.  The resulting image is
a wavelength averaged field distribution in the waveguide as the
source is a broadband LED. Examination of the field pattern
reveals the dramatic change from a nearly Gaussian mode confined
predominantly in the slab center for the properly terminated
waveguide ($W_a$=0.8$a$) to a broadened mode that is delocalized
over the whole slab owing to the excitation of surface states when
$W_a$=1.2$a$.

\begin{figure}[tb]
\begin{center}
\leavevmode \psfig{figure=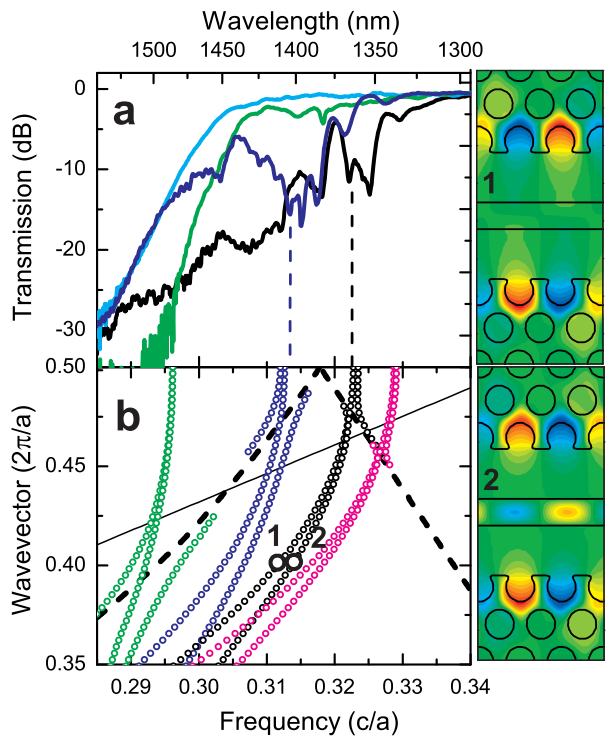,width=85mm}
\end{center}
\caption{(a): Set of transmission spectra for TE-polarized light
for DT PhC waveguides with different widths of the trench. Spectra
were measured with low resolution (5nm) on waveguides 500$\mu$m
long. The cyan, green, blue and black spectra correspond to the
samples with the trench width 1.6$a$, 1.0$a$, and two different
nominally 1.2$a$, respectively. (b) Set of photonic band
structures calculated for the TE-like modes for the DT PhC
waveguides with different trench widths.  Green, blue, black, and
magenta open circles shows dispersion of only the surface modes
for the trench width of 1.0$a$, 1.12$a$, 1.17$a$, and 1.2$a$,
respectively. The dashed black curve represents the undisturbed
dispersion of the PhC waveguide when surface states are pushed out
of the photonic gap (Wa=0.8$a$). The thin solid black line
represents the light line of the silica cladding layer. Right hand
panel: In-plane magnetic field profiles of the two
surface-localized modes calculated for $W_a$=1.12$a$ for the
reduced wavevector of 0.4 (numbered 1 and 2 and marked in the main
figure by open black circles).}
\label{fig3}
\end{figure}

To further analyze the interaction of the waveguiding mode with
surface states, photonic band structures were calculated
\cite{ref9}. Different trench widths $W_a$ and hence different
termination of the PhC surface are examined. As a reference the
dispersion of the waveguiding mode for trench width of 1.6$a$ is
shown in Fig.\ \ref{fig3}b by a dashed black line. It is almost
identical to the dispersion of the corresponding isolated strip
waveguide, which is close to linear for most of the frequency
range of interest.  The sharp cut-off seen around 1440nm in the
spectrum for the $W_a$ =1.6$a$ (green line in Fig.\ \ref{fig3}a),
is consistent with the mode crossing the light line set by the
silica cladding at 0.304$c/a$.

The photonic band diagram for the PhC with the trench width
$W_a$=1.2$a$ is shown by the magenta open circles in Fig.\ \
ref{fig3}b. It is almost identical to the previous case except for
the appearance of two surface modes which cross the fundamental
mode at 0.327$c/a$. For comparison with the band diagrams the
experimental transmission spectra are presented in Fig.\
\ref{fig3}a for $W_a$=1.6$a$, $W_a$=1.0$a$ and two nominally
designed $W_a$=1.2$a$ but fabricated with different exposure
conditions.  The origin of the sharp dips in the experimental
spectra can be associated with the crossing of the surface modes
with the guiding mode in the photonic band diagram. The sharp dip
in the experimental spectrum for $W_a$=1.2$a$ (shown by a black
line in Fig.\ \ref{fig3}a appears at 0.323$c/a$, a frequency
slightly lower than the crossing point of 0.327$c/a$ seen in the
band diagram. This roughly 3\% difference could reasonably be
attributed to a 3\% error in determination of the lattice
parameters of the experimental structure. Indeed if a slightly
smaller trench width of $W_a$=1.17$a$ is assumed the photonic band
calculations show a crossing at 0.323$c/a$ (black open circles in
Fig.\ \ref{fig3}b). The sharp dips around 1400nm in the blue
spectrum in Fig.\ \ref{fig3}a (also a nominally $W_a$=1.2$a$
device but with a lower exposure dose) can be fitted in an
analogous way by the band diagram for a trench width $W_a$=1.12$a$
(blue open circles in Fig.\ \ref{fig3}b). The transmission
spectrum for a PhC with a nominal trench width of $W_a$=1.0$a$
(green line in Fig.\ \ref{fig3}a does not differ considerably from
the reference spectrum for $W_a$=1.6$a$ (cyan line in Fig.\
\ref{fig3}a). However close inspection reveals enhanced
attenuation at frequencies below the light line cut-off, where
bands associated with surface states cross the fundamental mode at
around 0.295$c/a$.

\begin{figure}[tb]
\begin{center}
\leavevmode \psfig{figure=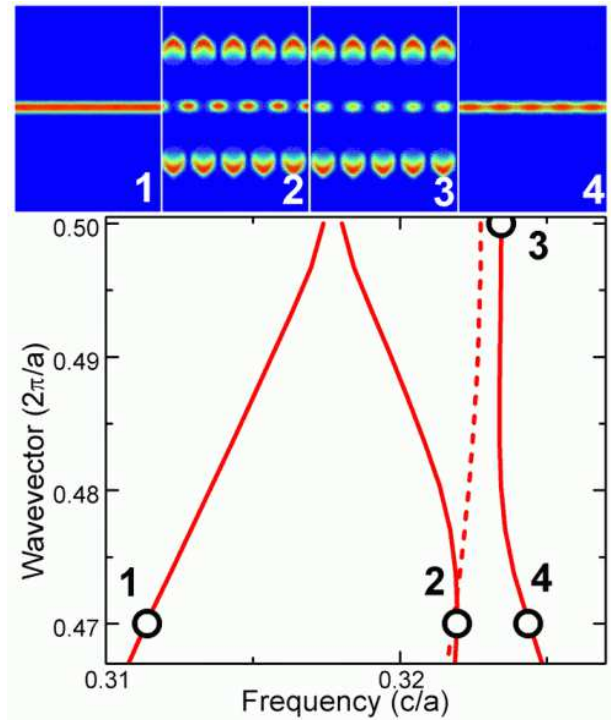,width=85mm}
\end{center}
\caption{Magnified view of the photonic band diagram for the
TE-like modes for the DT PhC waveguides with the trench width of
1.17$a$. Solid (dashed) lines correspond to bands of even (odd)
symmetry with respect to the $xz$ plane bisecting the slab along
the waveguide direction. Images 1-4 represent the in-plane
time-averaged magnetic field power density calculated for
corresponding wavevectors marked 1-4 in the photonic band
diagram.}
\label{fig4}
\end{figure}

Further insight into the coupling to surface states can be
provided by inspection of the magnetic field profiles. The right
hand panel of Fig.\ \ref{fig3} represents field profiles for the
case of $W_a$=1.17$a$ calculated for the reduced wavevector of 0.4
far from the crossing region. Two surface modes form a pair of
even and odd states with respect to the $xz$ plane bisecting the
slab along the waveguide direction. The odd mode (point 1) can not
be excited from the input access strip waveguide and,
correspondingly, it crosses the strip-localized fundamental mode
without interaction and does not contribute to transmission. The
even mode (point 2) has the same symmetry as the fundamental mode
in the strip and, correspondingly, they can interact with each
other at the crossing. This interaction can be further facilitated
by the considerable field intensity of the even surface mode in
the central strip waveguide.

Figure \ \ref{fig4}, a magnified portion of the band diagram for
$W_a$=1.17$a$ around 0.323$c/a$, reveals classical anti-crossing
behavior. The time-averaged magnetic field energy density profiles
for the fundamental mode related to the strip waveguide and the
even surface mode are presented on top of Fig.\ \ref{fig4}. It is
seen that for frequencies far away from the anti-crossing region
(point 1) the field profile is localized predominantly in the
central strip. At the anti-crossing of the strip-related mode and
even surface mode two mixed states are formed: one which is mostly
surface localized (point 2) and another mostly strip-localized
(point 4). A mini-stopband forms at the anti-crossing resulting in
strong attenuation of the transmission spectrum.

The transmission dip has a pronounced doublet structure indicating
a more complex interaction. It can be suggested that the spectral
shape of the transmission spectrum in the anti-crossing region is
strongly affected by the efficiency of coupling from the access
strip waveguide to the DT PhC waveguide. It is seen that the edges
of the mini-stopband are formed by two states (points 2 and 3 in
Fig.\ \ref{fig4}), which are predominantly surface-localized, and,
correspondingly, have poor coupling efficiency to the access strip
waveguides. Therefore attenuation can occur even for frequencies
beyond the boundaries of the mini-stopband, with the spectral
shape dependent on the way in which the energy in the mode is
redistributed between states predominantly localized in the strip
and the surface.

In conclusion, the influence of surface termination on the
dispersion of surface states in truncated PhCs is experimentally
demonstrated in a DT PhC waveguide system. Using photonic band
calculations it is shown that sharp dips in the transmission
spectra originate from coupling to surface modes.  It is
demonstrated that depending on the exact location of the
truncation with respect to the unit cell of the PhC lattice the
surface states can either be positioned at the center of the
photonic gap where they interact strongly with the waveguiding
mode in the strip waveguide, or be pushed away from the gap and do
not contribute to transmission.

\end{document}